\documentclass[iop,twocolumn]{emulateapj}
\bibpunct[; ]{(}{)}{;}{a}{}{;}
\usepackage{apjfonts}

\usepackage{graphicx} 

\newcommand{\Msolar}{M${_\odot}$\,}

\shorttitle{Star formation in NGC\,346 across time and space}
\shortauthors{De Marchi, Panagia \& Sabbi}

\slugcomment{Accepted for publication in ``The Astrophysical Journal''}

\begin{document}

\title{Clues to the star formation in NGC\,346 across time and
space\,\altaffilmark{*}}


\author{
Guido De Marchi,\altaffilmark{1}
Nino Panagia,\altaffilmark{2,3,4} 
and Elena Sabbi\altaffilmark{2}
}

\altaffiltext{1}{European Space Agency, Space Science Department,
Keplerlaan 1, 2200 AG Noordwijk, Netherlands; gdemarchi@rssd.esa.int}

\altaffiltext{2}{Space Telescope Science Institute, 3700 San Martin
Drive,  Baltimore, MD 21218, USA, panagia@stsci.edu, sabbi@stsci.edu}

\altaffiltext{3}{INAF--CT, Osservatorio Astrofisico di Catania, Via S.
Sofia  78, 95123 Catania, Italy}

\altaffiltext{4}{Supernova Limited, OYV \#131, Northsound Rd., Virgin
Gorda,  British Virgin Islands}

\altaffiltext{{$\star$}}{Based on observations with the NASA/ESA
{\it Hubble Space Telescope}, obtained at the Space Telescope Science
Institute, which is operated by AURA, Inc., under NASA contract
NAS5-26555}

\begin{abstract}  

We have studied the properties of the stellar populations in the field
of the NGC\,346 cluster in the Small Magellanic Cloud, using the results
of a novel self-consistent method that provides a reliable
identification of pre-main  sequence (PMS) objects actively undergoing
mass accretion, regardless of their age. The 680 identified bona-fide
PMS stars show a bimodal age distribution, with two roughly equally
numerous populations peaked respectively at $\sim 1$\,Myr, and $\sim
20$\,Myr. We use the age and other physical properties of these PMS
stars to study how star formation has proceeded across time and space in
NGC\,346. We find no correlation between the locations of young and old
PMS stars, nor do we find a correspondence between the positions of
young PMS stars and those of massive OB stars of similar age.
Furthermore, the mass distribution of stars with similar age shows large
variations throughout the region. We conclude that, while on a global
scale it makes sense to talk about an initial mass function, this
concept is not meaningful for individual star-forming regions. An
interesting implication of the separation between regions where massive
stars and low-mass objects appear to form is that high-mass stars might
not be ``perfect'' indicators of star formation and hence a large number
of low-mass stars formed elsewhere might have so far remained unnoticed.
For certain low surface density galaxies this way of preferential 
low-mass star formation may be the predominant mechanism, with the
consequence that their total mass as derived from the luminosity may be 
severely underestimated and that their evolution is not correctly
understood.

\end{abstract}


\section{Introduction}

With over 30 O-type stars amongst its denizens (Massey, Parker \&
Garmany 1989; Evans et al. 2006), the NGC\,346 cluster is the site of
most intense star formation in the Small Magellanic Cloud (SMC) as well
as one of  the most active in the Local Group. The massive young stars
in NGC\,346 are responsible for the ionisation of the surrounding N\,66
nebula, the largest HII region in the SMC (Henize 1956). The location of
NGC\,346 and N\,66 in the SMC, their geometry and the limited amount of
foreground extinction have made these regions an ideal place to study the 
effects of massive objects on the surrounding medium, including whether 
they can effectively trigger the formation of new generations of stars, 
as some theories of sequential star formation suggest (e.g. Elmegreen \&
Lada 1977). 

Over the past 20 years, many authors have attempted to give an answer to
these questions. Massey, Parker \& Garmany (1989) conducted a
photometric and spectroscopic study of NGC\,346, revealing not only 33
O-type stars ({\small $1/3$} of which earlier than O$6.5$), but also
several lower-mass stars ($\sim 15$\,\Msolar) forming a distinct
subgroup $\sim 2\farcm6$ to the SW of the centre and with an  estimated
age of order 15\,Myr. The age difference with respect to the O-type
stars led these authors to suggest that sequential star formation might
have occurred in the region. 

Using near- and mid-infrared as well as CO sub-millimetre observations,
Contursi et al. (2000) and Rubio et al. (2000) were able to identify
several embedded sources in the bar making up the body of NGC\,346,
corresponding to strong emission peaks whose presence may reveal recent
and/or ongoing star formation. However, while the peak corresponding
with the central NGC\,346 cluster contains unreddened stars, all other
peaks are affected by higher reddening, suggesting that the interstellar
material has not been completely ejected and that they could be in a
younger stage of evolution. Since the more reddened peaks appear to be
located farther away from the central cluster, these authors have
suggested that star formation might have taken place in a sequential way
along the bar. 

More recent observations with the Hubble and Spitzer Space Telescopes 
have resolved the mid-infrared emission peaks, revealing that they are
compact clusters made up of a multitude of pre-main sequence (PMS) stars
(Nota et al. 2006; Sabbi et al. 2007; Hennekemper et al. 2008) and young
stellar objects (YSO; Bolatto et al. 2007; Simon et al. 2007). { 
Hennekemper et al. (2008) performed a detailed analysis of the locations
of these PMS stars in the colour--magnitude diagram (CMD), which they
compare with the PMS isochrones of Siess et al. (2000) taking into
account the effects of differential reddening, binarity and variability
on the age determination. They conclude that, depending on the amount of
reddening present in the field, the observed broadening of the positions
of these PMS objects in the CMD can be compatible with both a single
star formation episode some $\sim 10$\,Myr ago or with two episodes
about 5 and 10\,Myr ago.} Nevertheless, even in this latter case, the
lack of a correlation between the estimated ages and the positions of
the objects in the field led Hennekemper et al. (2008) to conclude that
there is no obvious signature of sequential star formation in this
region.  

{ In a subsequent study by the same team, Gouliermis et al. (2008)
suggest that signs of sequential star formation might actually be
present. They propose a scenario in which the birth of the three young
star clusters on the arc-like structure was triggered by the winds of
the massive progenitor of SNR\,B0057--724 located at the centre of the
arc,  $\sim 20$\,pc away. Gouliermis et al. (2008) also suggest that, in
a similar manner, the powerful winds of the OB stars at the centre of
NGC\,346 are shaping a dusty arc feature to the south and southwest of
the association. On the other hand, as we will show in Section\,3, the
dusty arc feature appears to be $\sim 20$\,Myr old and little or not at
all affected by the presence of the OB stars in the centre (Smith
2008), suggesting that there is no causal connection between the arc and
the massive stars at the centre of NGC\,346.}

The weak side of all these studies is that they are mostly qualitative.
They are primarily based on the analysis of the spatial distribution of
the objects and on the morphology and geometry of the features present
in the field. However, they do not take into account the actual ages of 
the { low-mass} stars that are needed to establish whether there are
dependencies and correlations amongst the stellar generations that have
formed in the recent past in these regions. 

{ An improvement in this sense is offered by the recent work of
Cignoni et al. (2011). Using a classical synthetic CMD procedure, they
concluded that NGC\,346 has experienced different regimes of star
formation, including a dominant and focused ``high density mode'', which
according to these authors led to the formation of rich and massive
sub-clusters hosting both PMS and massive MS stars, and a subsequent
diffuse ``low density mode'', characterised by the presence of
sub-clusters hosting PMS stars only. These different modes of star
formation can have an impact on the shape of the mass function (MF), as
we discuss further in Section\,3 and 4. Cignoni et al. (2011) suggest
that the richest sub-clusters formed $\sim 6$\,Myr ago, with an apparent
remarkable synchronisation, while star formation in the sub-clusters
mainly composed of PMS stars appears to have started $\sim 3$\,Myr ago,
following a multi-seeded spatial pattern. 

Even in works of this type, however, the available age range for
low-mass stars is limited to the youngest PMS objects ($\lesssim
5$\,Myr). } This is because ages are based on the comparison of the
observations with theoretical evolutionary tracks in the
Hertzsprung--Russell (H--R) diagram, and at older ages the isochrones
become too close to the very populous main sequence (MS) of field stars
to provide reliable results. 

Actually, the presence of distinctive emission features in the  spectra
of PMS stars with ages up to $\sim 30$\,Myr, due to the accretion
process to which they undergo, allows us to efficiently and securely
detect and identify all objects of this type in a stellar field,
regardless of their age and of their position in the H--R diagram.
Building on the work of Romaniello (1998) and of Panagia et al. (2000),
De Marchi, Panagia \& Romaniello (2010, hereafter Paper\,I) showed that
through a suitable combination of broad- and narrow-band photometry it
is also possible to derive the mass accretion rate of these objects,
with an accuracy comparable to that allowed by spectroscopy. In a
companion paper (De Marchi et al. 2011; hereafter Paper\,II), we
applied the method developed in Paper\,I to the high-quality HST
photometry of NGC\,346 (Sabbi et al. 2007) and were able to identify two
distinct generations of bona-fide PMS stars (about 700 objects) with a
clearly bimodal age distribution in the range from $\lesssim 1$\,Myr to
$\sim 30$\,Myr. In this work we use the accurate physical parameters
that we have measured in Paper\,II and correlate them  with the spatial
distribution of these objects. The availability of accurate ages for
such a large number of stars across the field is the key element that
was missing in previous studies of NGC\,346 and allows us for the first
time to study how star formation has proceeded in this area over the
past $\sim 30$\,Myr. 

The paper is organised as follows: in Section\,2 we briefly summarise
the results of Paper\,II and presents the relevant observational
material.  Section\,3 compares the spatial distribution of the two
generations of PMS  stars to one another and to that of young massive
stars present in the field. In Section\,4 we look at how the shape of
the stellar mass  function changes across the field. In Section\,5 we
discuss the possible consequences of different processes operating for
high mass and low mass star formation on the study of galaxies and the
determination of their star formation rates. A summary of the most
important conclusions of the paper is offered in Section\,6.

\section{Pre-main sequence stars in NGC\,346}

In Paper\,II, we have studied the properties of the stellar populations 
in a field $200\arcsec \times 200\arcsec$ around the centre of NGC\,346,
making use of observations collected with the Advanced Camera for
Surveys on board the Hubble Space Telescope (details on the observations
and on the photometric analysis of the data can be found in Nota et al.
2006 and Sabbi et al. 2007). We refer the reader to Paper\,II for a
detailed description of the analysis of the stellar populations in this
field and of the determination of their properties. However, for
convenience, we offer hereafter a brief summary of the main results that
are most relevant to this paper. 

Thanks to a novel, self-consistent method  developed in Paper\,I, it is
possible to reliably identify PMS stars undergoing active mass
accretion, regardless of their age. The method, fully described in
Paper\,I and II, does not require spectroscopy and combines broad-band
$V$ and $I$ photometry with narrow-band $H\alpha$ imaging to detect all
stars with excess H$\alpha$ emission while simultaneously providing an
accurate measure of their accretion luminosities $L_{\rm acc}$ and mass
accretion rates $\dot M_{\rm acc}$. 

The application of this method to the NGC\,346 observations allowed us
to reveal 791 PMS candidates, namely objects with H$\alpha$ excess above
the $4\,\sigma$ level with respect to the reference provided by normal
cluster stars observed in the same bands. The average H$\alpha$
luminosity of these PMS candidates is $2.7 \times 
10^{31}$\,erg\,s$^{-1}$ or $\sim 10^{-2}$\,L$_\odot$. In order to avoid
possible contamination due to objects with significant chromospheric
activity, we retained as bona-fide PMS stars only those with a large
equivalent width of the H$\alpha$ emission line, $W_{\rm eq} <
-20$\,\AA\  for stars with $T_{\rm eff} < 10\,000$\,K or $W_{\rm eq}
<-50$\,\AA\ for hotter stars (note that, as customary, a negative
equivalent width is used for emission lines). A total of 694 objects
satisfy these conditions.

\begin{figure}[t]
\centering
\resizebox{\hsize}{!}{\includegraphics[width=16cm]{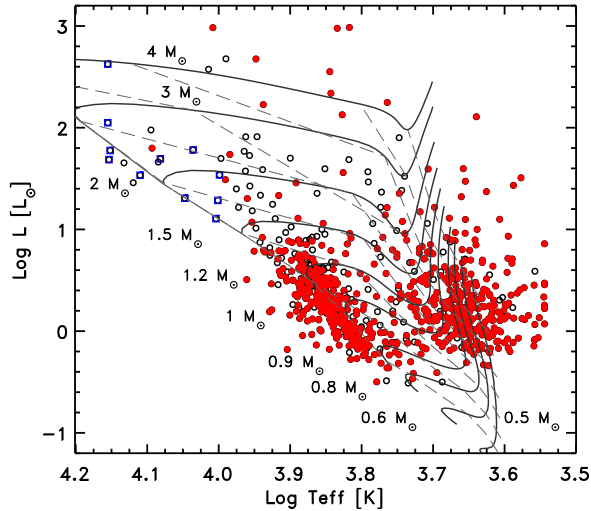}}
\caption{Hertzsprung--Russell diagram of the PMS candidate stars. All
objects shown here have excess emission in H$\alpha$ at the $4\,\sigma$
level or higher. Those indicated with filled circles also have  $W_{\rm
eq}(H\alpha) < -20$\,\AA\ or $< -50$\,\AA\ for stars hotter than
10\,000\,K. Squares correspond to objects with $W_{\rm eq}(H\alpha) > -
50$\,\AA\ and, as such, are potential Be stars. Thick solid lines show
the evolutionary tracks from Degl'Innocenti et al.  (2008) for
metallicity $Z=0.002$ and masses from $0.5$ to 4\,\Msolar, as indicated.
The corresponding isochrones are shown as thin lines, for ages of
$0.125$, $0.25$, $0.5$, 1, 2, 4, 8, 16 and 32\,Myr from right to left.
Note that the constant logarithmic age step has been selected in such a
way that the photometric uncertainties are smaller than the distance
between  the isochrones in the H--R diagram. } 
\label{fig1}
\end{figure}

{ A colour--magnitude diagram showing the positions of these
objects in the observational plane ($V$ vs $V-I$) is provided in
Paper\,II.} In Figure\,\ref{fig1} we show the locations of these objects
in the H--R  diagram (thick dots, in red in the online version),
compared with the PMS evolutionary models of the Pisa group
(Degl'Innocenti et al. 2008; Tognelli, Prada Moroni \& Degl'Innocenti
2011) for metallicity $Z=0.002$. As noted in Paper\,II, although this
metallicity is at the lower end of the currently accepted values for the
SMC, ranging from $\sim {\small 1/5}$ to $\sim {\small 1/8}$\,Z$_\odot$
(see Russell \& Dopita 1992; Rolleston et al. 1999; Lee et al. 2005;
Perez--Montero \& Diaz 2005), it appears better suited to describe the
properties of young PMS stars in this field. As Figure\,\ref{fig1}
immediately shows, there are two separate groups of objects, occupying
two distinct regions in the diagram, one well above (or redder than) the
MS and the other at the MS itself. { This distinction is equally
clear in the CMD shown in Paper\,II}.

Comparison with the evolutionary tracks (solid lines, corresponding to
the masses as indicated) and with the isochrones (dashed lines, for ages
increasing from right to left from $0.125$\,Myr to 32\,Myr doubling at
each step) allows us to determine accurate masses and relative ages for
680 of these objects. Note that previous determinations of these
parameters for candidate PMS stars in NGC\,346 that made use of
evolutionary models for $Z=0.01$ are necessarily less accurate. { For
the interpolation, we followed the procedure developed by Romaniello
(1998), which does not make assumptions on the properties of the
population, such as the functional form of the initial mass function
(IMF). On the basis of the measurement errors, this procedure provides
the probability distribution for each individual star to have a given
value of the mass and age (the method is conceptually identical to the
one presented recently by Da Rio et al. 2010a).}

The masses that we obtain range from $\sim 0.4$\,\Msolar to $\sim
4$\,\Msolar, with an average value of $\sim 1$\,\Msolar, whereas the
ages show a clear bimodal distribution, { already implicit in
Figure\,\ref{fig1}}, where very few objects are seen around ages of
$\sim 4-8$\,Myr. Taking 7\,Myr as a threshold, the PMS stars can be
split into two almost equally populous groups. One group comprises 
350 objects younger than 7\,Myr (hereafter called ``younger PMS
stars''), with a median age of $\sim 1$\,Myr (formally $0.9$\,Myr) and a
distribution ranging from $0.3$\,Myr to $3.1$\,Myr, respectively the 17
and 83 percentiles. The other group contains 330 objects older than
7\,Myr (hereafter called ``older PMS stars''), with a median age of
$\sim 20$\,Myr (formally $19.7$\,Myr) and a distribution ranging from
$12.5$\,Myr to $26.5$\,Myr (17 and 83 percentiles, respectively). {
In fact, the latter value should be considered a lower limit to the age
of these objects. Regardless of the statistical approach that one
follows, when a star in the H--R diagram is closer to the zero-age 
MS than its photometric errors, only a lower limit to its age can be
derived, since the long tail of the distribution function makes all
older ages in principle equally likely.}

As discussed in Paper\,I, our age uncertainties are typically less than
a factor of two (hence the choice of the spacing between isochrones in
Figure\,\ref{fig1}) { and this includes systematic differences
arising because of the use of models that might not properly describe
the stellar population under study (e.g. because of the wrong
metallicity) and from differences between models of various authors. Yet
for a given set of models, our photometric uncertainties result in even
smaller uncertainties on the relative ages, typically of order $\sqrt{2}$.
Thus,} the age difference {\em between} the two groups of PMS stars is
very significant, { since the age gap between the two groups is much
wider than the uncertainty on the relative ages}. In fact, the age
spreads {\em within} each group might actually be somewhat smaller than
what we quote, since our interpolation procedure does not take
unresolved binaries into account. Da Rio et al. (2010a) have shown that
when binaries are ignored the derived age spread appears wider, although
the average age itself is not affected. Therefore, the two groups of
objects may be even better separated in age than our numbers imply. On
the other hand, we believe that interpreting the observed broadening as
an age effect is the most conservative assumption as regards our
conclusions. We address the properties of these two populations in
detail in Section\,3.

{ Besides the age and mass and H$\alpha$ luminosity already
mentioned, another other important physical parameter that we derived in
Paper\,II for these objects is the mass accretion rate $\dot M_{\rm
acc}$, which has a median value of $3.9 \times
10^{-8}$\,\Msolar\,yr$^{-1}$. This value is about 50\,\% higher than
that measured in Paper\,I for a population of 133 PMS stars in the field
of SN\,1987A, owing to the much younger median age of PMS objects in
NGC\,346. In fact, the large size of our PMS sample and its spread in
mass have allowed us to study the evolution of the mass accretion rate
as a function of stellar parameters and to conclude that $\log \dot
M_{\rm acc} \simeq -0.6 \log t + \log m + c$, where $t$ is the age of
the star, $m$ its mass and $c$ a quantity that is higher at lower
metallicity (see Paper\,II for details). }

\section{Star formation across time and space}

In this section we use the information on the physical parameters of
PMS stars of various masses and ages to study how star formation has
proceeded in NGC\,346 over the past $\sim 30$\,Myr, since this is the
time span that we can effectively and accurately probe with our method.

\subsection{Multiple stellar generations}

As mentioned above, the distribution of our bona-fide PMS stars in
the H--R diagram reveals a shortage of objects with ages around $\sim
4-8$\,Myr, suggesting a likely gap or lull in star formation at that
time. The presence of two so clearly distinct groups of stars with
H$\alpha$ excess can only be interpreted as the result of distinct star
formation episodes. Hillenbrand et al. (2008) and Hillenbrand (2009)
have argued that random luminosity spreads apparent in the H--R diagram
of star forming regions and young clusters are often erroneously
interpreted as true luminosity spreads and taken as indicative of true
age spreads. This is clearly not the case here, since no random spread
in the luminosity of a single age population could produce such a
distinctive bimodal distribution in the H--R diagram.

{ On the other hand, some of the objects that we label older PMS
stars could actually be very young stars with a circumstellar disc seen
at high inclination ($> 80^\circ$). Objects of this type would appear  
bluer than their photospheric colour due to light scattering on the 
circumstellar disc. However, they would also be several magnitudes
fainter than their photospheric brightness due to extinction caused by
an almost edge-on disc. According to the models of Robitaille et al.
(2006) for the spectral energy distribution of young stars seen at
various viewing angles, objects of this type can only account for a few
percent of the total young population. Furthermore, as we will show in
Section\,3.2, the spatial distribution of the stars with H$\alpha$
excess close to the MS is remarkably different from that of the younger
PMS stars, and this should not be the case if these were all objects of
the same type simply viewed at different inclinations. Therefore the
vast majority of stars with H$\alpha$ excess near the MS must be
intrinsically older. 

In a forthcoming paper (De Marchi, Guarcello \& Panagia, in preparation)
we will address in detail the role played by circumstellar discs seen at
high inclination, which can undoubtedly account for a small fraction ($<
5$\,\%) of our sample. That work will discuss in detail the theoretical
implications that such a geometry can have on the extinction and
scattering of the light of the central object and will address the
specific case of NGC\,6611 in the Eagle Nebula, where like in NGC\,346 a
population of older ($\sim 10$\,Myr) PMS stars is also present. }

\begin{figure}[t]
\centering
{\includegraphics[scale=0.5,bb=54 180 558 562]{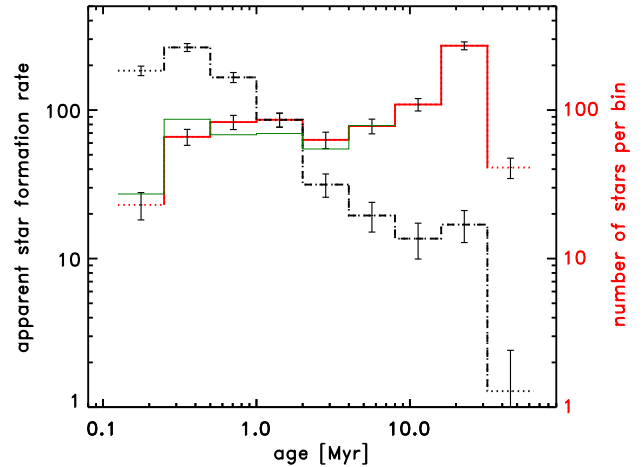}}
\caption{Histograms showing the number of stars per age bin (thick solid
line) and the apparent star formation rate (dot-dashed line) as a
function of age. Only bona-fide PMS stars (i.e. objects with H$\alpha$
excess emission at the $4\,\sigma$ level or above at the time of the
observations) in the range $0.4 - 4$\,\Msolar are considered in this
figure, so the dot-dashed line provides a lower limit to the true star
formation rate. The thin solid line shows the age distribution of all
stars (i.e. also those without H$\alpha$ excess emission), but only up
to ages of 8\,Myr, since older objects cannot be distinguished from
field MS stars. The thin solid histogram is shifted vertically by
$-0.56$\,dex and appears in excellent agreement with the thick solid
histogram.}
\label{fig2}
\end{figure}

As regards the age distribution of PMS stars in NGC\,346, a histogram is
shown in Figure\,\ref{fig2}, where PMS ages are binned using a constant
logarithmic step (a factor of 2) that better reflects the relative age
uncertainties stemming from the comparison of model isochrones with the
actual data. The solid line in Figure\,\ref{fig2} gives the number of
stars inside each age bin as a function of time, whereas the dot-dashed
line provides an apparent value of the star formation rate, in units of
stars per Myr, derived by dividing the number of objects in each bin by
the width of the bin. Note that at the extremes of the distribution it
becomes more difficult to assign an age to the stars, thus the first and
last bin are drawn with a dotted line to indicate a larger uncertainty.
{ In particular, as mentioned above, for stars that in the HR diagram
are closer to the MS than their photometric error, the age that we
provide is in practice a lower limit to the true age. This is due to the
fact that the distribution function is characterised by such an extended
tail towards older ages that all ages older than the value that we
provide are virtually equally likely.  

Also the dot-dashed line necessarily represents a lower limit to the star 
formation rate. In this case, the reason is not the age uncertainty but
the fact that the we consider exclusively} the number of detected PMS stars
in the range $0.4 - 4.0$\,\Msolar that at the time of the observations
had H$\alpha$ excess emission at the $4\,\sigma$ level or above. One
limitation is caused by photometric incompleteness at low masses, which
makes it more difficult to detect faint PMS stars in crowded
environments. Another effect is the uncertainty on the fraction of PMS
stars that at any given time show excess H$\alpha$ emission. For younger
PMS stars ($\lesssim 8$\,Myr), whose position in the H--R diagram is
well separated from that of field MS stars, this fraction can be
estimated from the ratio of stars with and without H$\alpha$ excess in
the same region of the diagram. The data show that at the time of the
observations this ratio was $0.28 \pm 0.04$ (see also Paper\,II). 

\begin{figure*}[t]
\centering
\includegraphics[scale=0.7]{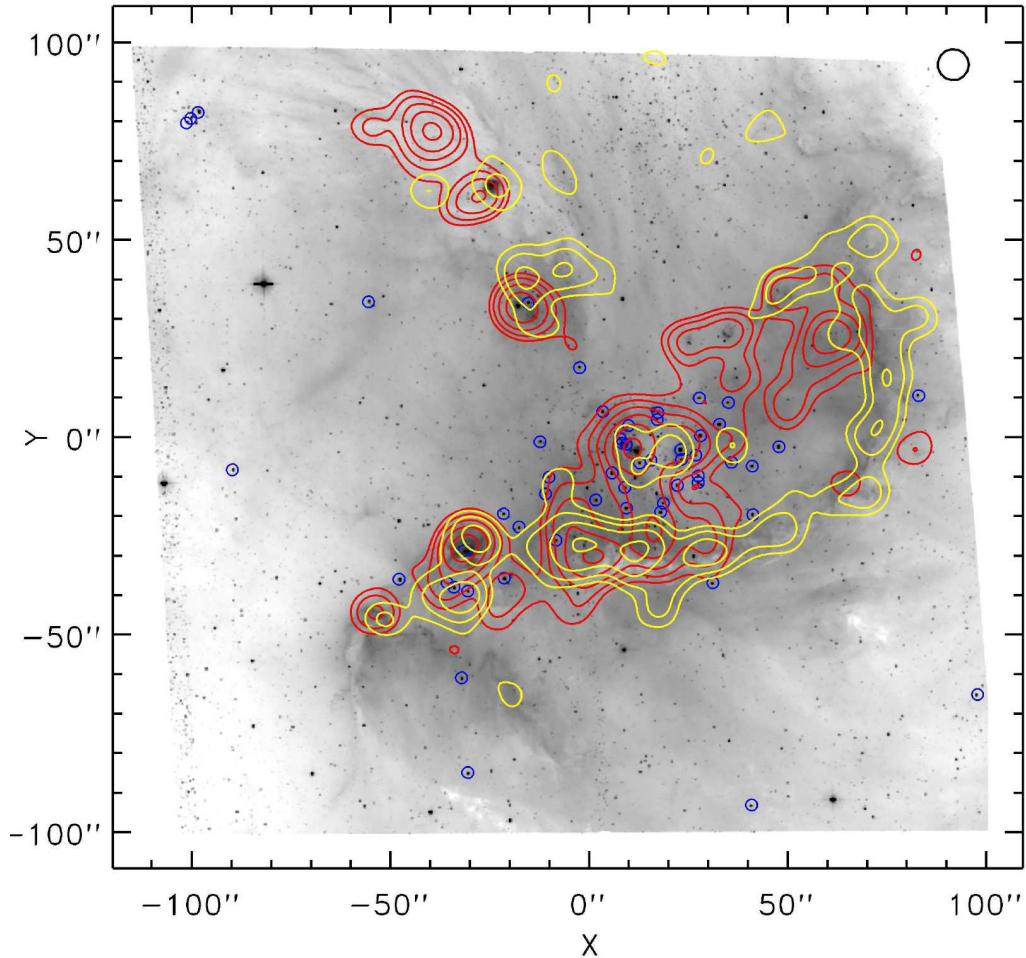}
\caption{The contour lines show the spatial density distribution of
young ($< 7$\,Myr; red) and old ($> 7$\,Myr; yellow) PMS stars in
NGC\,346, overlaid on a negative H$\alpha$ image of the region. The (0,0)
position in this figure corresponds to RA$=0^{\rm h} 59^{\rm m} 8^{\rm
s}$, DEC$=-72^\circ 10\arcmin 32\arcsec$ (J2000), while North is up and
East to the left. The contour plots have been obtained after Gaussian
smoothing with a beam size of $\sigma=4\arcsec$, as indicated by the
circle in the upper right corner. Blue circles correspond to young ($<
7$\,Myr) massive stars brighter than $\sim 2 \times 10^4$\,L$_\odot$.
The contour levels have a logarithmic spacing of $\sim 0.17$\,dex (or a
factor of $1.5$). The lowest level corresponds to a density of $0.033$
stars per arcsec$^2$, equivalent to twice the average density of PMS
stars in this field. The highest contour level for old PMS stars
corresponds to $0.11$ stars per arcsec$^2$ and that for young PMS stars
to $0.25$ stars per arcsec$^2$.}
\label{fig3}
\end{figure*}

The thin solid line in Figure\,\ref{fig2} (green in the online version)
shows the age distribution of all stars in the H--R diagram younger than
8\,Myr, shifted vertically by  $-0.56$\,dex { (corresponding to a
factor of $0.28$),} and is in excellent agreement with the histogram of
bona-fide PMS stars. However, it is presently not known how this ratio 
would change at older ages, and it is also expected to depend on the mass 
of the stars, so at this stage it is only possible to set a lower limit 
to the true star formation rate. Near the peak of the distribution, at 
$\sim 0.4$\,Myr, this limit corresponds to $\sim 200$\,\Msolar\,Myr$^{-1}$ 
(the median mass of those objects is $\sim 0.7$\,\Msolar), while at $\sim 
25$\,Myr it drops by an order of magnitude to 
$\sim 20$\,\Msolar\,Myr$^{-1}$ (median mass $\sim 1$\,\Msolar). 

At face value, the star formation strength for stars in the range
$0.4-4.0$\,\Msolar thus appears to be much higher in the present burst
than in the one { that was active $\sim 10 - 30$\,Myr ago and might 
have ended $\sim 8$\,Myr ago,} while the total integrated output (i.e.
the total number of stars, as shown by the thick solid histogram) in the
two episodes is rather similar. However, there are important selection
effects that one must consider. As mentioned above, one effect is the
fraction of PMS stars with H$\alpha$ excess, which most likely varies
throughout the PMS phase, making it more difficult to compare the number
of younger and older PMS stars to one another. Another problem is that,
while the previous burst has certainly ended, the current one might
still continue for a long time, thus making it hard to predict how many
stars will eventually be formed. Finally, the accuracy on relative ages
being at best of order a factor of $\sqrt{2}$, it is not possible to say
exactly how long the previous burst lasted nor how many short bursts
took place in the time frame covered by one age bin. This carries the
implication that, if there was just one short burst, the star formation
strength might have been comparable to or even higher than that of the
current episode. 


The results of our quantitative age analysis on the star formation
efficiency in this field are still necessarily tentative, but it will be
possible to lift at least some of the uncertainties plaguing this
picture through a systematic comparison of different star forming 
clusters in similar states of evolution that we plan to conduct in the 
future. On the other hand, having established that there are at least
two star formation episodes in NGC\,346, separated by $\sim 10$\,Myr or
more, we still can study whether they are independent of one another or
appear to be causally connected.

\subsection{Looking for spatial correlations: contour plots}

{ We compare in Figure\,\ref{fig3} the spatial density distributions
of the younger and older PMS populations by means of contour lines with
logarithmic scaling, overlaid on a H$\alpha$ image of NGC\,346. As
mentioned above, the two groups include respectively 350 and 330
objects.} The contour plots have been obtained after smoothing the
distribution with a Gaussian beam with size $\sigma=4\arcsec$ or $\sim
1.2$\,pc, as indicated by the circle in the upper right corner of the
figure. The lowest contour level corresponds to a local density of PMS
stars twice as high as the average PMS stars density over the entire
field. The step between contour levels is constant and corresponds to a
factor of $1.5$. We also show with small circles the positions of 55
young massive stars brighter than $\sim 2 \times 10^4$\,L$_\odot$ and
with an implied mass $> 15$\,\Msolar (e.g. Iben 1967).

A striking feature in this figure is the difference in the spatial
distribution of the three types of stars. Many older PMS stars are
distributed along the rim of the gas shell to the S and W of the
cluster's centre (hereafter named ``southern arc'') and, except for the
centre itself, they appear to avoid regions where younger PMS stars are
located. As for massive stars, albeit more abundant near the centre of
NGC\,346, they also appear at various other locations in the field that
are not occupied by PMS objects.

{ It is interesting to compare these contour lines with the maps of
mid-IR emission obtained with ISOCAM on board the Infrared Space
Observatory by Contursi et al. (2000) and Rubio et al. (2000). The
strong emission peaks discovered by these authors (see also
Introduction) coincide with regions of recent star formation in our
analysis as well, and in particular with the intensity peaks due to
younger PMS stars in Figure\,\ref{fig3} (darker contour plots, shown in
red in the online version). As regards older PMS stars, including those
in the southern arc, they appear projected against a background of lower
IR emission. From the analysis of observations with the Spitzer Space
Telescope, Simon et al. (2007) discovered in the arc a few young stellar
objects of relatively high mass ($> 4.5$\,\Msolar). As we will show
later (see Figure\,\ref{fig4}), there are indeed also some younger PMS
stars along this gas rim, but the arc appears to be mostly dominated by
older PMS objects.}

The careful reader could be worried that several old PMS stars seem to
lie along the southern arc, since nebular emission might in principle
contaminate their photometry and give us an inaccurate measurement of
their H$\alpha$ excess emission. However, as already mentioned in
Paper\,II, we have carefully inspected the images and removed from the
list of bona-fide PMS stars all objects whose H$\alpha$ photometry might
be contaminated by gas filaments. While it is possible that some of the
H$\alpha$ emission that we detect is due to diffuse nebular emission in
the HII region not powered by the accretion process (see Paper\,I), if
the emission is extended and uniform over an area comparable to that of
the point spread function, its contribution cancels out with the rest of
the background when we perform the  photometry (see Sabbi et al. 2007
and Paper\,II for details on the photometry). 

Obviously, the subtraction would not work if the emission were not
uniform, as for example in the case of a filament that projects over the
star but that does not cover completely the background annulus. For this
reason, after applying an unsharp-masking algorithm to highlight and
sharpen the details of the H$\alpha$ frames, we have carefully inspected
all sources with excess H$\alpha$ emission and have marked as suspicious
and excluded from our bona-fide sample all those with filaments
contamination within $0\farcs3$ of the star, for a total of 62 objects.
Although some of them might have intrinsic H$\alpha$ excess emission, we
prefer to adopt a conservative approach and remove all dubious cases. A
detailed example of how well this powerful technique works can be found
in Beccari et al. (2010). Therefore, we are confident that the tight
distribution of older PMS stars along the rim of the gas shell is not an
artefact and suggests instead that these objects have very low
velocities or at least a very small velocity spread. 

Observed values of the velocity dispersions of stars in young clusters
and associations typically fall in the range $1 - 10$\,km\,s$^{-1}$
(e.g. van Altena et al. 1988; Jones \& Walker 1988; Mengel et al. 2009;
Bosch, Terlevich \& Terlevich 2009; Rochau et al. 2010). The thickness
of the projected distribution of the older PMS in NGC\,346 is of order
$\sim 20\arcsec$ or $\sim 6$\,pc. With a median estimated age of $\sim
20$\,Myr, this implies a small value of the projected velocity spread,
namely $\la 0.5$\,km\,s$^{-1}$, corresponding to a three dimensional
velocity dispersion of $\la 1$\,km\,s$^{-1}$ along the gaseous rim. This
picture is consistent with the very low velocity dispersion ($< 3$\,km
s$^{-1}$) of the ionised gas measured by Smith (2008) in this field from
high-resolution echelle spectroscopy. If there is a higher velocity
component, it must be linked to the systematic motion of the gas shell.

The match between the location of many old PMS objects (about $\sim$
{\small $1/3$} of them) and the rim of the gas shell also suggests that
the shell itself reflects the distribution of the gas out of which these
stars formed and that it has not (yet) been significantly affected by
the stellar winds and by the ionising radiation of the much younger
massive stars at the centre of the field. Furthermore, the fact that the
distribution of these massive objects and of the younger PMS stars does
not appear to trace in any way the geometry of the gas shell indicates
quite convincingly that we are seeing two rather different and unrelated
generations of stars.

From the apparent shape of the rim of the gas shell, Gouliermis et al.
(2008) recently argued that there is a relationship between the central
NGC\,346 cluster and the southern arc. They suggested that the
latter outlines the ionisation front of  the cloud that is caused by the
powerful stellar winds of the young massive stars at its centre. In
their scenario, the photoionisation process of the central OB stars
would provide the primary source of mechanical energy that triggers star
formation in this region. { Our analysis does not support this
interpretation:} not only is the rim of the gas shell unaffected by the
central OB stars, but it is also much older (and it may be much farther
away from the OB stars than what the projected distance might seem to
suggest). This discrepancy outlines the risks of drawing conclusions on
triggered star formation based primarily on the morphology of structures
projected on the sky. As Watson, Hanspal \& Mengistu (2010) have
recently shown, only 20\,\% of the sample of HII regions that they
studied appear to have a significant number of YSOs associated with
their photodisociation fronts, implying that triggered star formation
mechanisms acting on the boundary of the expanding HII region are not
common. 


\subsection{Looking for correlations: number ratios}

\begin{figure*}[t]
\centering
\resizebox{\hsize}{!}{\includegraphics{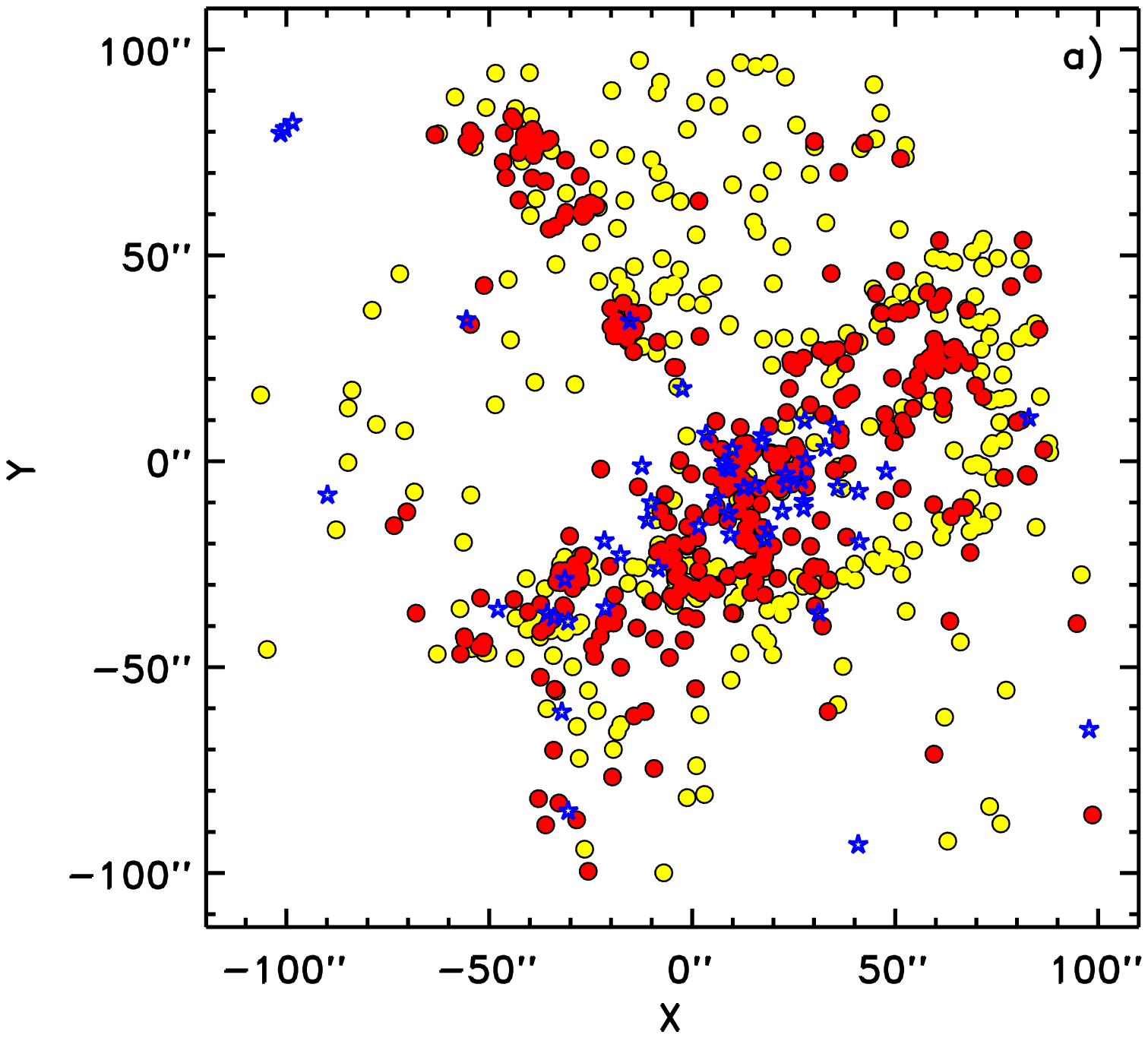}
\includegraphics{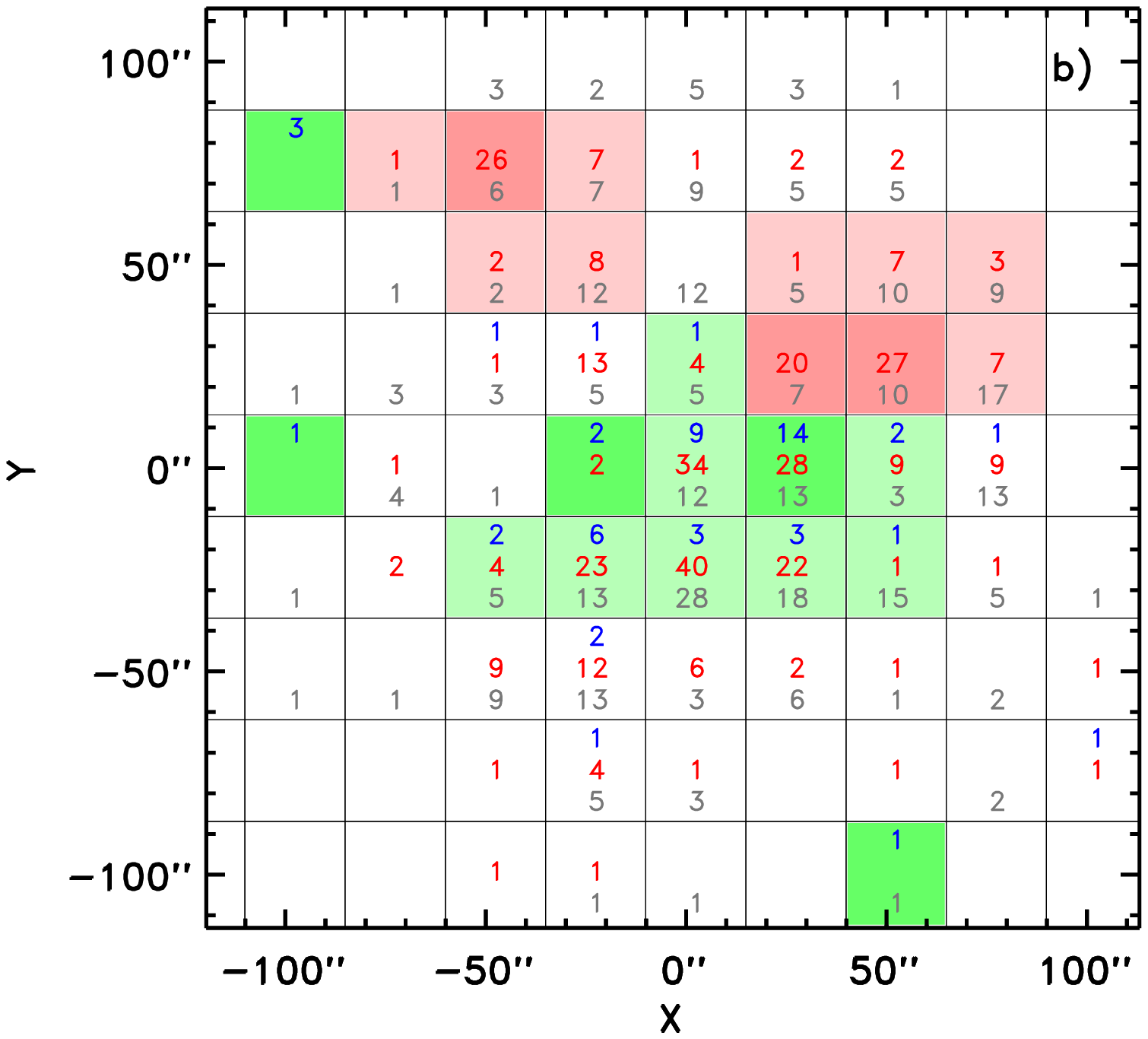}}
\caption{Panel a): relative locations of young PMS stars (darker
circles, red in the online version), old PMS stars (lighter circles,
yellow in the online version) and  massive MS stars (pentagrams, blue in
the online version). Panel b): after placing a uniformly spaced grid on
panel a), we have counted the number of stars of each type falling into
cells of $25\arcsec$ on a side. The corresponding counts are listed in
each cell, with the number of massive stars on top, the number of young
PMS stars in the middle and that of older PMS object at the bottom of
each cell. See text for the meaning of the background colours of the
cells. }
\label{fig4}
\end{figure*}

A more quantitative characterisation of the relative distribution of
younger and older PMS objects and massive stars is offered by the maps
shown in Figure\,\ref{fig4}. The (0,0) position in that figure
corresponds to RA$=0^{\rm h} 59^{\rm m} 8^{\rm s}$, DEC$=-72^\circ
10\arcmin 32\arcsec$ (J2000), with North up and East to the left. In
panel a) we show all young stars using different symbols (blue
pentagrams, red dots and yellow dots respectively for massive young
stars, younger PMS stars and older PMS objects), whereas panel b) gives
the number of stars of each type falling within cells of $25\arcsec$ or
$7.3$\,pc on a side, which is the typical size of a star cluster in the
Magellanic Clouds  ($7.7 \pm 1.5$\,pc; Hodge 1988).

The difference in the distribution of younger and older PMS stars
already seen in Figure\,\ref{fig3} continues to be present in
Figure\,\ref{fig4}, revealing that older PMS objects are less
concentrated and more widely distributed than younger stars. The other
notable feature in Figure\,\ref{fig4} (and already visible in
Figure\,\ref{fig3}) is the mismatch between the positions of young
massive stars and those of young PMS objects of similar age (both types
of objects are younger than 7\,Myr and most are not older than 3\,Myr).
The majority of young massive stars are clustered near the adopted
centre of NGC\,346, where a high concentration of young PMS stars is
also seen. However, there is also a number of massive objects that are
not surrounded by an overdensity of PMS stars. Similarly, many PMS
objects are clustered in populous groups with no massive stars in their
vicinity, although the total mass of the groups can reach $\sim
100$\,\Msolar even when only considering PMS objects in the range $0.4 -
4$\,\Msolar, { once photometric completeness is taken into account.}
A similar situation is seen in the Cygnus X North complex (Beerer et al.
2010), where the positions of many early B-type stars (the most massive
objects in the region) do not coincide with those of young star clusters
of similar age.

These differences can be easily quantified by using the star numbers
shown in the cells of Figure\,\ref{fig4}b. The numbers provide the count
of massive stars on top, of young PMS stars in the middle and of older
PMS object at the bottom of each cell. Since in this field there are a
total of 55 massive stars ($> 15$\,\Msolar) and 350 young PMS stars, on
average one would normally expect in each cell about 6 times more young
PMS objects than massive stars of similar age, but as the figure shows
this is not always the case. On this basis, assuming Poisson statistics,
we note at least two regions with many PMS stars where the observed
absence of massive stars has a probability of less than 5\,\% to occur
by chance. They are centered around ($+60\arcsec$, $+25\arcsec$) and
($-40\arcsec$, $+70\arcsec$). This condition is true for the individual
cells with darker shade (darker pink in the online version), while for
the light-shaded cells in their vicinities (lighter pink in the online
version) the condition still applies if they are combined with
neighbouring cells of the same or darker colour. 

This simple statistical test proves that the observed paucity of massive
young stars in these two rather wide areas is significant at a
$2\,\sigma$ level, at least. { In fact, since our photometric
completeness is worse in the most central regions (see Sabbi et al.
2007), the paucity of massive young stars is even more pronounced there.
A very similar result has been recently found by Cignoni et al. (2011),
who analysed the ratio of massive MS stars and low-mass PMS objects
within the individual sub-clusters detected by Sabbi et al. (2007) in
this region. Although the identification of PMS objects in the work of
Cignoni et al. (2011) is purely based on their broad-band colours and
magnitudes, and as such can in principle be affected by interlopers and
field objects, these authors conclude that the PMS stars in the regions
corresponding to our pink cells in Figure\,\ref{fig4}b are
over-represented with respect to massive stars, for a typical IMF.}

Similarly, one can find in Figure\,\ref{fig4}b several cells (shaded in
darker green in the online version) where the number of young PMS
objects falls below the 5\,\% Poisson probability expected from the
number of massive stars. { Some of these cells are located at the
centre of NGC\,346, where crowding and photometric incompleteness make
it more difficult to detect fainter PMS stars, so the statistical
significance of our non detection of low-mass stars is lower. However, 
most of the isolated massive stars are located in the outer regions of 
the cluster, where photometric completeness and crowding are not a
concern}. It is possible that these objects are not cluster members, but
the fact that they are not surrounded by PMS stars of similar age
remains puzzling and might imply that they have been displaced from
their formation region due to dynamical interactions, such as in the
case of star 30\,Dor~016 (Evans et al. 2010).

A possibility would be that these massive objects were the lower-mass
companions of disrupted binary systems in which the primaries have
already ended their evolution. The radial velocity study of NGC\,346 by
Evans et al. (2006) suggests that at least {\small $1/4$} of the massive
stars are in binary systems. The typical projected separation of the
isolated massive stars from the centre of NGC\,346 is of order 2\arcmin\
or $\sim 35$\,pc. Given the young ages prevalent in NGC\,346 (less than
3\,Myr), none of the youngest generation stars could have produced
runaways. Very few isolated massive stars may indeed be runaways from
binary systems if they were part of a $\sim 15$\,Myr old population,
such as the one associated with the subcluster SC--16 (Sabbi et al.
2007) and corresponding to the three massive objects in the upper left
corner of Figure\,\ref{fig3} at ($-100\arcsec$, $+80\arcsec$). In this
case, runaway stars could cross the entire field of, say, 2\arcmin\ or
$\sim 35$\,pc moving at a velocity of about 3\,km\,s$^{-1}$ for about
10\,Myr. 

This however seems hardly to be the case for the objects marked by
circles in that figure. Firstly, none of them has an age in excess of
7\,Myr, according to our photometry and to the spectroscopy of Evans et
al. (2006) for the objects in common with their catalogue. Secondly, the
fact that there are only three massive stars in the direct vicinity of
the $\sim 15$\,Myr cluster would require that most massive stars were in
binary systems and that almost all secondaries have moved away from it:
this explanation appears rather contrived and seems to require a very
unlikely occurrence. Therefore, we have to accept that the lack of
massive stars associated with a number of PMS star clusters simply
reflects a genuine property of star formation in those parts of the the
NGC\,346 complex.

\begin{figure*}[t]
\centering
{\includegraphics[scale=0.4]{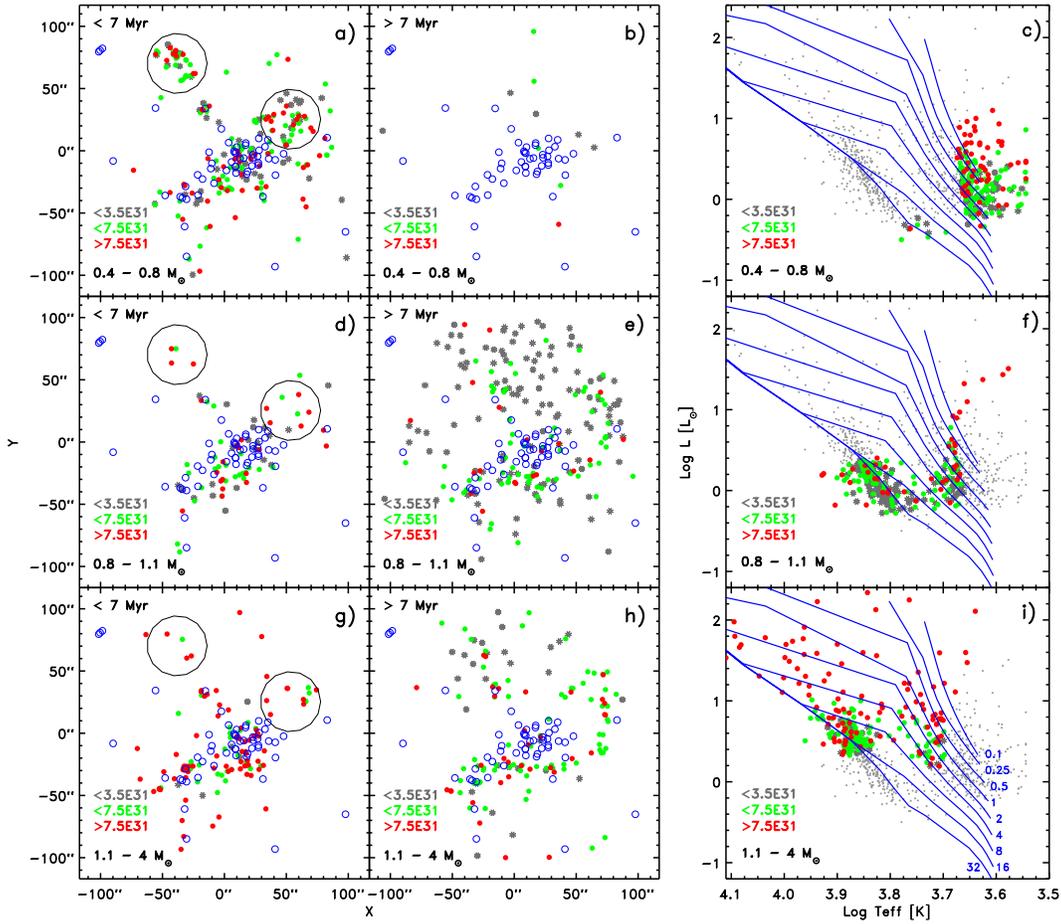}}
\caption{Spatial distribution and position in the H--R diagram of all
our bona-fide PMS stars, as a function of their age, mass and
$L(H\alpha)$ luminosity. Objects are sorted according to their age ($<
7$\,Myr or $> 7$\,Myr), to their mass ( $0.4-0.8$\,\Msolar,
$0.8-1.1$\,\Msolar and  $1.1 - 4$\,\Msolar) and to their H$\alpha$
luminosity ($< 3.5\times10^{31}$ erg s$^{-1}$, $< 7.5\times10^{31}$ erg
s$^{-1}$ and $> 7.5\times10^{31}$ erg s$^{-1}$, respectively grey
asterisks, green dots and red dots in the online version) as per the
legends shown in each panel. The small circles (blue in the online
version) indicate  the positions of the 55 young massive stars already
shown in Figures\,\ref{fig3} and \ref{fig4}a. The two large circles in
Panels a), d) and g) refer to the regions described in the text. As for
the H--R diagrams, the solid lines correspond to PMS isochrones from the
Pisa group for metallicity  $Z=0.002$ and ages in Myr as indicated in
panel i) with a constant  logarithmic step (factor of 2). The small dots
correspond to stars with masses outside of the range indicated in each
panel.}
\label{fig5}
\end{figure*} 

In summary, from the analysis of Figure\,\ref{fig4} we can conclude in a
more quantitative way that there is little or no spatial correlation
between the position of young massive objects and that of PMS stars of
similar age, except for the centre of NGC\,346, and that the two
populations of older and younger PMS objects have rather different
distributions (except again possibly for the regions near the cluster's
centre). This implies that the stellar MF varies considerably across the
field, as Cignoni et al. (2011) have recently suggested. Therefore, if
the MF is sampled over a limited region, its shape will not be
statistically representative of the entire cluster. As we discuss in the
next section, this has profound implications for the concept of IMF.

\section{Uniformity of the mass function}

Before proceeding to study the MF of PMS stars, we must consider the
possible selection effects inherent in our measurements and how they
could affect our analysis of the MF. For this purpose we summarise most
of the relevant information in Figure\,\ref{fig5}, to which we will
refer throughout this section. Shown in the figure are the spatial
distributions and positions in the H--R diagram of all our bona-fide PMS
stars, as a function of their ages, masses and L(H$\alpha$) luminosities
as measured in Paper\,II. As before, we split objects into two age
groups, younger or older than 7\,Myr, while for the mass we use three
bins, namely $0.4-0.8$\,\Msolar, $0.8-1.1$\,\Msolar and $1.1 -
4$\,\Msolar. We also use symbols of different colours (most easily
discernible  in the online version of the paper) to identify stars with
different H$\alpha$ luminosities, as indicated by the legends. In all
panels the small open circles refer to the positions of the same 55
young massive stars already shown in Figures\,\ref{fig3} and
\ref{fig4}a. As for the H--R diagrams, the solid lines correspond to PMS
isochrones from the Pisa group (Degl'Innocenti et al. 2008; Tognelli et
al. 2011) for metallicity $Z=0.002$ and ages as indicated in panel i)
with a constant logarithmic step ($0.3$\,dex corresponding to a factor
of two in age). Since the photometric uncertainty is typically smaller
than the separation between neighbouring isochrones, { with the 
interpolation procedure explained in Section\,2 we are able to assign 
relative ages with an accuracy of better than a factor of two, typically
of order $\sqrt{2}$.}

A first obvious selection effect, which will however not affect our
determination of the MF of young PMS stars, is revealed by the paucity
of PMS objects in the range $0.4 - 0.8$\,\Msolar and age $> 7$\,Myr.
This is due to the detection limit of our photometry at $\sim
0.5$\,L$_\odot$.

Another interesting characteristic apparent from this figure, and
already partly seen in Figures\,\ref{fig3} and \ref{fig4}b, is the
compact distribution of older PMS stars of all masses along the gaseous
rim of the southern arc. The fact that the majority of these objects are
rather luminous ($L(H\alpha) > 7.5 \times 10^{31}$\,erg s$^{-1}$)
confirms that they are not artefacts due to the gas rim itself, as
already discussed in Section\,3.

To illustrate the extent of the MF variations present in this field, we
start from panel a), where the two conspicuous groups of young low-mass
($0.4 - 0.8$\,\Msolar) PMS objects already seen in Figure\,\ref{fig4}
are visible around ($+60\arcsec$, $+25\arcsec$) and ($-40\arcsec$,
$+40\arcsec$), as indicated by the large circles with a radius of $\sim
25\arcsec$ or $\sim 7.3$\,pc. Interestingly, very few objects are found
in the same regions in panels d) and g), where equally young stars with
masses of respectively $0.8 - 1.1$\,\Msolar and $1.1 - 4$\,\Msolar are
shown. In particular, in panel a) there are 36 and 41 PMS stars in these
regions, whereas 4 and 8 are found in panel d) and 4 and 9 in panel g),
respectively. Based on the number of stars inside the circles in panel
a) and assuming a power-law MF of the type $dN/dm \propto m^\alpha$ with
$\alpha=-2.0 \pm 0.2 $, as typically observed in Galactic star forming
regions (De Marchi, Paresce \& Portegies Zwart 2010), one would expect
respectively 11 and 13 objects inside the circles in panel d) and 12 and
14 in panel g). These values are considerably larger than the numbers
observed, particularly for the region at ($+60\arcsec$, $+25\arcsec$).
If we were to derive the slope of the MF in the mass range $0.4 -
4$\,\Msolar from the ratio of the number of stars occupying the same
region in the three  panels, we would obtain a rather steep, MF with
$\alpha = -3.1 \pm 0.3$ (respectively $-2.8 \pm 0.3$ in the first region
and $-3.6 \pm 0.4$ in the second). Note that the true MF slope is most
likely even steeper, owing to the fact that our smaller mass bin is
subject to some incompleteness, as mentioned above, even though the
completeness of our photometry is well above 50\,\% for all but a
handful of PMS stars (see Paper\,II).


Although instructive, the comparison between panels a), d) and g)
carried out in this way is subject to some bias. We are using the
relative number of stars in different mass bins to characterise the
shape of the MF, but the objects that we consider are only those
undergoing active mass accretion as witnessed by excess H$\alpha$
emission. { This necessarily introduces some selection effects
because more massive stars reach the MS more quickly than lower-mass
objects (e.g. Palla \& Stahler 1993; see also Paper\,II) and it becomes
increasingly more difficult to derive an accurate age for them. If these
objects are associated with older ages, they could be systematically
underrepresented in our sample, thereby causing a spurious steepening of
the MF. }

A possible way to avoid these effects would be to limit the mass range
for the analysis, i.e. to consider only panels a) and d) since none of
the objects in the mass range $0.4 - 1.1$\,\Msolar is expected to have
reached the MS at an age of $<7$\,Myr. Alternatively, one could limit
the age range and only consider very young stars, e.g. those with $<
1$\,Myr, which are all in the PMS phase over the entire mass range
considered here ($0.4 - 4$\,\Msolar). This second approach has the
advantage of providing more specific information on the IMF and we will
therefore follow it.

Interestingly, however, even when restricting our analysis to stars
younger than 1\,Myr we still find a very steep MF: there are in total 35
objects with this age in the two reference areas of panel a), but only
two objects are found in each of the corresponding regions in panel d)
and g). The implied MF slope in the range $0.4 - 4$\,\Msolar remains
very steep, { namely $\alpha = -4.3 \pm 0.5$}. One could argue that, if 
the areas that we are studying are too small, the steep MF slopes that we
obtain might be an artefact of small numbers statistics. However, with a
projected radius of $25\arcsec$\ or $\sim 7.3$\,pc, the circles in
panels a), d) and g) are twice as large as the typical size of star
forming clusters in the Magellanic Clouds (e.g. Hodge 1988). Therefore,
these regions sample a sufficiently large area to be representative of
the local conditions of star formation, that in this case appear to be
characterised by the the paucity of massive stars and correspondingly a
rather steep MF slope. A similarly steep MF ($\alpha=-4 \pm 0.5$) was
found  by Massey (2002) in his study of the field of the Magellanic
Clouds, i.e. in regions similarly devoid of massive stars.

\begin{figure}[t]
\centering
{\includegraphics[scale=0.5,bb=54 180 558 600]{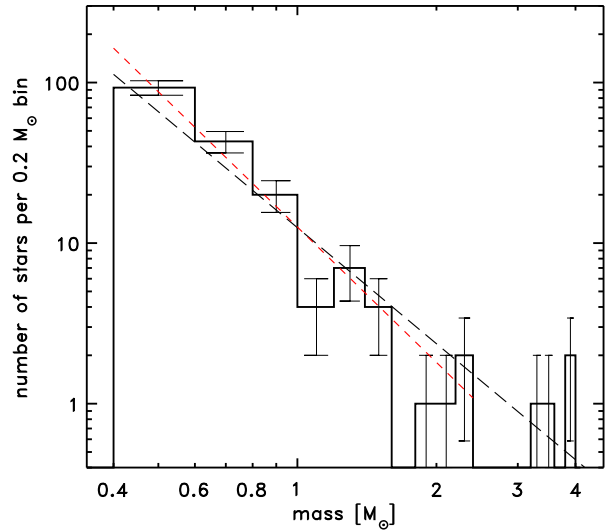}}
\caption{Mass function of PMS stars with age $< 1$\,Myr. The
short-dashed line is the best power-law fit in the mass range
$0.4-2$\,\Msolar and corresponds to $\alpha = -2.8 \pm 0.2$. The
long-dashed line is the fit over the entire mass range, with $\alpha =
-2.4 \pm 0.4$\,\Msolar. }
\label{fig6}
\end{figure} 

A more ``typical'' value of the MF slope for stars younger than 1\,Myr
could be derived if we extended our analysis to the entire area covered
by these observations, corresponding to $\sim 60$\,pc on a side. In
their study of the NGC\,346 region, Sabbi et al. (2008) determined the
present day MF in NGC\,346 from the same observations that we use in
this paper. When considering all stars in the field in the range $0.8 -
60$\,\Msolar, they derived a MF slope $\alpha = -2.43 \pm 0.18$, very
close the the Salpeter (1955) value. When we consider all PMS stars
younger than 1\,Myr in this field { we find a good match with the
results of Sabbi et al. (2008) over the common mass range, namely $
0.4 - 4$\,\Msolar. This is shown in Figure\,\ref{fig6}, where the 
long-dashed line corresponds to $\alpha = -2.43$ and provides a good fit
to the mass distribution of PMS stars younger than 1\,Myr. However, we
still find a somewhat steeper slope ($\alpha = -2.8 \pm 0.2$,
short-dashed line) in the mass range $0.4 - 2$\,\Msolar, where the
statistics is more robust thanks to the larger number of objects.}

In reality, the true MF slope must be slightly shallower than the
$\alpha$ values that we have obtained since we only consider as
bona-fide PMS stars those showing H$\alpha$ excess emission at the
$4\,\sigma$ level or above. As discussed in Paper\,I and II, the
accretion process and with it the H$\alpha$ luminosity is subject to
large variations (a factor of 2 -- 3 in a few days; e.g. Fernandez et
al. 1995; Smith et al. 1999; Alencar et al. 2001), with the implication
that, statistically, not all PMS stars in a given region will show at
the same time H$\alpha$ excess emission { above our conservative
acceptance threshold}. If the fraction of PMS stars with H$\alpha$
excess emission were constant, considering only these objects would not
affect the determination of the  MF slope. However, as we will show in a
forthcoming paper (De Marchi et al., in preparation), that fraction
appears to become smaller when mass and age increase. While age is not
an issue in the present case, since we are only considering stars
younger than 1\,Myr, our objects span about a decade in mass ($0.4 -
4$\,\Msolar) and in this range variations as large as a factor of 3 are
seen in the fraction of PMS stars with H$\alpha$ excess emission. If
this effect were taken into account, the slope of the MF over the entire
field in the range $0.4 - 2$\,\Msolar would drop slightly, to
$\alpha=-2.0 \pm 0.3$, a value in line with the typical MF slopes and
corresponding uncertainties that are observed in nearby Galactic star
forming regions (e.g. De Marchi, Paresce \& Portegies Zwart 2010;
Bastian, Covey \& Meyer 2010).

Nevertheless, even applying this correction to the MFs measured in the
two regions discussed above (see the large circles in
Figure\,\ref{fig5}) would still give considerably steeper indices
($\alpha \simeq -3$) than those commonly measured in star clusters.
Therefore, the general conclusion  that we can draw from this analysis
is that there are considerable variations in the shape of the MF of
young PMS stars across the field of NGC\,346, as witnessed by largely
different values of the power-law index $\alpha$. These variations
remain even when we only consider the MF of stars younger than 1\,Myr,
which would normally be taken as representative of the IMF.

Since these variations are seen when comparing groups of objects
comprising several low-mass stars distributed over the typical size of a
stellar cluster ($7.5$\,pc radius), they cannot be ascribed to simple
statistical fluctuations. Coupled with the remarkable anti-correlation
between massive and low-mass stars of similar age (see Section\,3 and
Figure\,\ref{fig4}), this indicates quite convincingly that the
formation of high- and low-mass stars requires at least different
initial conditions, and might also be governed by different mechanisms.
{ This is also the conclusion recently reached by Cignoni et al.
(2011). From the observation that the apparently youngest sub-clusters,
i.e. those composed only by stars in the PMS phase, show a deficiency 
of massive stars, these authors speculate that the IMF may be a function
of time, with the youngest sub-clusters not having had sufficient time
yet to form more massive objects.} As already pointed out by Panagia et
al. (2000), while the concept of IMF might be meaningful over large
areas where it represents the average result of different star formation
processes, its predictive power over smaller scales, characteristic of a
specific stellar cluster or association, will have to be seriously
reconsidered.

\section{Discussion}

An interesting corollary of the apparent separation between massive star
formation sites and regions where moderate/low mass stars form is that,
in general, massive stars may turn out to not be ``perfect'' indicators
of active star formation in galaxies. While it is true that regions
where massive stars have formed are actively forming stars, it is not
necessarily true that those regions identify the places where most of
the stars are formed. In fact, integrating over a Salpeter IMF ($\alpha
=-2.35$) defined over the range $0.1-150$\,\Msolar it is easy to
calculate that objects above 4\,\Msolar represent only about 20\% of the
total mass that goes into stars. The fraction is even lower in the field
of the Magellanic clouds, where Massey (2002) found $\alpha=-4 \pm 0.5$.

Our analysis of the  NGC\,346 observations, as well other fields in the
LMC and SMC (e.g. Panagia et al. 2000; Romaniello, Robberto \& Panagia
2004; De Marchi, Romaniello \& Panagia 2010; and in preparation)
indicates that low mass stars formation appear to run independently of
the formation of massive stars. This may be due to the fact that even in
a big cloud that is subdivided into smaller sub-units, massive stars can
only form inside the larger cells whereas low-mass stars may form
readily in smaller cells. If the bulk of the mass in the  interstellar
medium is not in the form of massive clouds, it is possible that the
formation of low mass stars be quite active in regions which are not
marked by the presence of massive stars. It is also possible that under
suitable conditions this separate channel of low mass star formation be
the dominant process to form stars.  

Within this scenario, it is clear that measuring the star formation
rates of external galaxies, especially in low surface density galaxies,
on the basis of massive star diagnostics (be it direct detection of OB
stars, optical and radio emission from HII regions, or their secondary
far-infrared radiation for highly opaque clouds), may indeed lead to
gross underestimates and to misleading results on the nature and the
evolution of galaxies.  

In order to clarify these issues one should intensively and
systematically study regions of local galaxies, selected in an unbiassed
way, where low mass stars can be individually detected and
characterized, so as to determine their physical parameters including
mass, luminosity, age and evolutionary status. This can be done not only
in our galaxy and, even better, in the Magellanic Clouds in which all
stars are approximately at the same distance from us, but also in many
other galaxies of the Local Group.  

For this purpose, in addition to a number of regions with obvious signs
of active star formation, such as the Orion Nebula Cluster (Da Rio et
al. 2010b, and in preparation) and NGC\,3603 in our Galaxy (Beccari et
al. 2010), 30\,Dor (De Marchi et al. 2011) and other selected fields in
the LMC, and NGC\,346 and NGC\,602 in the SMC, we are also analysing a
sample of fields selected from the Archival Pure Parallel
Project\footnote{See  http://archive.stsci.edu/prepds/appp for details.}
that were imaged with the HST-WFPC2 in four broad-band filters (F300W,
F450W, F606W, and F814W) and a narrow H$\alpha$ filter (F656N) with a
total exposure time of at least 3 orbits, i.e. about 160 minutes (Spezzi
et al. 2011). Although these fields are still not
completely ``random'' and ``unbiased'', since they were imaged in
parallel with long UV spectroscopic observations of OB stars taken with
another HST instrument about 5\arcmin\ to 12\arcmin\ away, they still
represent a relatively rich sample (about 20 fields in each LMC and SMC)
of regions with generally marginal massive star formation. For the
future it would be helpful to obtain complementary H$\alpha$
observations of HST archival fields with deep exposures in the $V$ and
$I$ bands, at least, as well as to target suitably selected new fields
in the Milky Way and in the Magellanic Clouds.

{ Finally, considerable improvements in this field will become
possible once PMS evolutionary models are properly calibrated. Even
though the models currently available are state of the art, they are not
as refined and extensively tested as those for MS stars or post-MS
evolution. In particular, absolute PMS ages are difficult to determine
due to the lack of independent calibrators. Improvements in these models
would make PMS stars even more powerful indicators of how star formation
proceeds over time and space.}

\section{Summary and conclusions}

We have studied the properties of the stellar populations in the field
of the NGC\,346 cluster in the Small Magellanic Cloud, using the results
of a novel self-consistent method that provides a reliable
identification of PMS objects actively undergoing mass accretion,
regardless of their age. We have used the age and other physical 
parameters measured for these PMS stars to study how star formation has 
proceeded across time and space in NGC\,346 over the past $\sim
30$\,Myr. The main results of this work can be summarised as follows.

\begin{enumerate}

\item 
The 680 identified bona-fide PMS stars show a bimodal age distribution,
with two roughly equally numerous populations with median ages of
respectively $\sim 1$\,Myr and $\sim 20$\,Myr, { although the latter
is most likely a lower limit to the true age. The age separation of the
two groups is consideraby wider than the uncertainties on the relative
ages} and the scarcity of objects with ages around $\sim 8$\,Myr
suggests a lull in star formation at that time. 

\item 
{ Although, taken at face value, the colours and magnitudes of the
older PMS stars are compatible with those of young PMS objects whose
light is absorbed and scattered by a high-inclination circumstellar
disc, this hypothesis is only viable for a few percent ($< 5$\,\%) of
the stars. This is confirmed by the remarkaby different spatial
distributions of the two age groups.} 

\item 
We set a lower limit to the star formation rate of the ongoing burst
of $\sim 200$\,\Msolar\,Myr$^{-1}$, while at $\sim 25$\,Myr the rate
drops by an order of magnitude to $\sim 20$\,\Msolar\,Myr$^{-1}$.
Both values are lower limits since they are based on the number of PMS
stars in the range $0.4 - 4$\,\Msolar that were undergoing active mass
accretion at the time of the observations.

\item 
At face value, the strength of the current star formation episode 
appears to be much higher than that of the one ended $\sim 8$\,Myr ago, 
while the total integrated output of the two episodes is rather similar.
However, since photometric uncertainty does not allow an age resolution
of better than a factor of $\sim \sqrt{2}$, we cannot establish how many
bursts took place between 8 and 30\,Myr ago, nor their duration. If
there was just one short burst, the star formation strength might have
been comparable to or even higher than that of the current episode.  

\item  
Except for the regions near the centre of NGC\,346, the stars belonging
to the two generations have markedly different spatial distributions. A
good fraction ($\sim$ {\small $1/3$}) of the older generation occupies
an arc-like gas structure to the south and west of NGC\,346 that had
been previously interpreted as the ionisation front caused by the OB
stars at its centre. { Although the morphology of the arc could have
suggested a case of triggered star formation, this is clearly not a
viable option} since the central massive stars are at least 10--20\,Myr
younger than the objects on the arc and cannot have triggered their
formation.

\item
The compact distribution of older PMS stars along the arc-like structure
suggests that they have formed there from the gas that is still visible
and have not (yet) been affected by the massive OB stars at the centre
of NGC\,346. This picture is consistent with the very low velocity
dispersion ($< 3$\,km s$^{-1}$) of the ionised gas measured in this
field from high-resolution echelle spectroscopy.

\item 
Except for the most central regions of NGC\,346, we find no
correspondence between the positions of young PMS stars and massive
O-type stars of similar age, suggesting that the conditions (and
possibly also the mechanisms) for their formation must be rather
different. Furthermore, the mass distribution of similarly aged stars 
shows large variations across the region. We conclude that, while on a
global scale it makes sense to talk about an initial mass function, this
concept is not meaningful for individual star forming clumps.

\item
It is possible that a large number of low-mass stars are forming in
regions where massive stars are not present and, therefore, they remain
unnoticed. For certain low surface density galaxies this might be the
predominant way of star formation, which would imply that their total 
mass based on the luminosity can be severely underestimated and that 
their evolution is not correctly understood. 

\end{enumerate}

\begin{acknowledgements}

We are grateful to an anonymous referee for extensive comments that have
helped us to improve the presentation of this work.  NP acknowledges
partial support by HST-NASA grants GO-11547.06A and GO-11653.12A, and
STScI-DDRF grant D0001.82435. 

\end{acknowledgements}


\begin{thebibliography}{References}
 

\bibitem[()]{} Alencar, S., Basri, G., Hartmann, L., Calvet, N. 
  2005, A\&A, 440, 595 

\bibitem[()]{} Bastian, N., Covey, K., Meyer, M. 2010, ARAA, 48, 339

\bibitem[()]{} Beccari, G., et al. 2010, ApJ, 720, 1108

\bibitem[()]{} Beerer, I., et al. 2010, ApJ, 720, 679 

\bibitem[()]{} Bolatto, A. D., et al. 2007, ApJ, 655, 212

\bibitem[()]{} Bosch, G., Terlevich, E., Terlevich, R. 2009, AJ, 137, 3437 

\bibitem[()]{} Cignoni, M., Tosi, M., Sabbi, E., Nota, A., Gallagher, J.  
  2011, AJ, 141, 31

\bibitem[()]{} Contursi, A., et al. 2000,  A\&A, 362, 310

\bibitem[()]{} Da Rio, N., Gouliermis, D. A., Gennaro, M. 2010a, ApJ, 723, 166

\bibitem[()]{} Da Rio, N., Robberto, M., Soderblom, D., Panagia, N., 
  Hillenbrand, L., Palla, F., Stassun, K. G. 2010b, ApJ, 722, 1092

\bibitem[()]{} Degl'Innocenti, S., Prada Moroni, P. G., Marconi, M., 
  Ruoppo, A. 2008, Ap\&SS, 316, 25

\bibitem[()]{} De Marchi, G., et al. 2011, ApJ, in press
  (arXiv:1106.2801)

\bibitem[()]{} De Marchi, G., Panagia, N., Romaniello, M. 2010, ApJ, 715,
  1 (Paper\,I) 

\bibitem[()]{} De Marchi, G., Panagia, N., Romaniello, M., Sabbi, E.,
  Sirianni, M., Prada Moroni, P.G., Degl'Innocenti, S. 2011, 
  ApJ, in press (arXiv:1104.4494, Paper\,II)

\bibitem[()]{} De Marchi, G., Paresce, F., Portegies Zwart, S. 2010, ApJ, 
  718, 105

\bibitem[()]{} Elmegreen, B., Lada, C. 1977, ApJ, 214, 725 

\bibitem[()]{} Evans, C., Lennon, D., Smartt, S., Trundle, C. 2006, A\&A, 
  456, 623 

\bibitem[()]{} Evans, C., et al. 2010, ApJ, 715, L74 

\bibitem[()]{} Fernandez, M., Ortiz, E., Eiroa, C., Miranda, L.  1995,
  A\&AS, 114, 439

\bibitem[()]{} Gouliermis, D., Chu, Y., Henning, T., Brandner, W.,
  Gruendl, R., Hennekemper, E., Hormuth, F. 2008, ApJ, 688, 1050

\bibitem[()]{} Gouliermis, D., Henning, T., Brandner, W., Dolphin, A.
  E., Rosa, M.,  Brandl, B. 2007, ApJ, 665, L27 

\bibitem[()]{} Henize, K. 1956, ApJS, 2, 315 

\bibitem[()]{} Hennekemper, E., Gouliermis, D., Henning, T., Brandner, W., 
  Dolphin, A. 2008, ApJ, 672, 914 

\bibitem[()]{} Hillenbrand, L. 2009, in The Ages of Stars, IAU Conf. Ser.
  258, eds. E. Mamajek, D. Soderbolm, R. Wyse (Cambridge: CUP), 81

\bibitem[()]{} Hillenbrand, L., Bauermeister, A., White, R., 2008, in 14th
  Cambridge Workshop on Cool Stars, Stellar Systems, and the Sun, ASP
  Conf, Ser. 384, ed. G. van Belle (San Francisco: ASP), 200

\bibitem[()]{} Hodge, P. 1988, PASP, 100, 1051 

\bibitem[()]{} Jones, B., Walker, M. 1988, AJ, 95, 1755 

\bibitem[()]{} Lee, J.-K., et al. 2005, A\&A, 429, 1025 

\bibitem[()]{} Massey, P. 2002, ApJS, 141, 81

\bibitem[()]{} Massey, P., Parker, J. W., Garmany, C. D. 1989, AJ, 98, 1305 

\bibitem[()]{} Mengel, S., Tacconi-Garman, L. 2009, Ap\&SS, 324, 321 

\bibitem[()]{} Nota, A., et al. 2006, ApJ, 640, L29

\bibitem[()]{} Palla, F., Stahler, S. 1993, ApJ, 418, 414 

\bibitem[()]{} Panagia, N., Romaniello, M., Scuderi, S., Kirshner, R. 2000,
  ApJ, 539, 197

\bibitem[()]{} P\'erez-Montero, E., D\'iaz, A. 2005, MNRAS, 361, 1063

\bibitem[()]{} Rochau, B., Brandner, W., Stolte, A., Gennaro, M., Gouliermis, 
  D., Da Rio, N., Dzyurkevich, N., Henning, T. 2010, ApJ, 716, L90 

\bibitem[()]{} Rolleston, W., et al. 1999, A\&A, 348, 728

\bibitem[()]{} Romaniello, M. 1998, PhD thesis, Scuola Normale Superiore,
  Pisa, Italy

\bibitem[()]{} Romaniello, M., Robberto, M., Panagia, N. 2004, ApJ, 608, 220 

\bibitem[()]{} Rubio, M., Contursi, A., Lequeux, J., Probst, R., Barba, R., 
  Boulanger, F., Cesarsky, D., Maoli, R. 2000, A\&A, 359, 1139

\bibitem[()]{} Russell, S., Dopita, M. 1992, ApJ, 384, 508 

\bibitem[()]{} Sabbi, E., et al. 2007, AJ, 133, 44 

\bibitem[()]{} Scuderi, S., Panagia, N., Gilmozzi, R., Challis, P.,
  Kirshner, R.  1996, ApJ, 465, 956 

\bibitem[()]{} Siess, L., Dufour, E., Forestini, M. 2000, A\&A, 358, 593

\bibitem[()]{} Simon, J., et al. 2007, ApJ, 669, 327

\bibitem[()]{} Smith, K., Lewis, G., Bonnell, I., Bunclark, P., Emerson, J. 
  1999, MNRAS, 304, 367

\bibitem[()]{} Smith, L. 2008, in Dynamical Evolution of Dense Stellar
  Systems, IAU Symp. 246, ed. E. Vesperini (Cambridge: CUP), 55

\bibitem[()]{} Spezzi, L., De Marchi, G., Panagia, N., Sicilia--Aguilar,
  A., Ercolano, B. 2011, MNRAS, submitted

\bibitem[()]{} Tognelli, E., Prada Moroni, P., Degl'Innocenti, S. 2011,
             A\&A, submitted

\bibitem[()]{} Watson, C., Hanspal, U., Mengistu, A. 2010, ApJ, 716, 1478

\bibitem[()]{} van Altena, W., Lee, J., Lee, J.-F., Lu, P., Upgren, A. 1988, 
  AJ, 95, 1744


\end{thebibliography}
\end{document}